\documentclass[prd,aps,preprint,superscriptaddress,showkeys,showpacs,
 preprintnumbers,amsmath,amssymb]{revtex4-1}
\usepackage{graphicx}
\usepackage{epsfig}
\usepackage{amsfonts}
\usepackage{dsfont}
\usepackage{amsmath}
\usepackage{amssymb}
\usepackage[sort&compress]{natbib}
\usepackage{dcolumn}
\usepackage{multirow}
\usepackage{bigstrut}
\usepackage{mathrsfs}
\usepackage{type1cm}
\usepackage{placeins}
\usepackage{color}
\usepackage{rotating}
\usepackage{slashed}
\usepackage{axodraw4j}
\usepackage{subfigure}
\usepackage[mathscr]{euscript}
\newcommand{\Xx}{\mathcal{X}}

\newcommand{\Hh}{\mathcal{H}}
\newcommand{\Ll}{\mathcal{L}}
\newcommand{\Mm}{\mathcal{M}}

\newcommand{\Ss}{\mathcal{S}}

\newcommand{\Aa}{\mathcal{A}}

\begin{document}

\title{CPT and Lorentz violation en the Photon and $Z$-boson sector}

\author{D. Colladay}
\affiliation{New College of Florida, Sarasota, FL, 34243, USA}

\author{J. P. Noordmans}
\author{R. Potting}
\affiliation{CENTRA, Departamento de F\'isica, Universidade do Algarve, 8005-139 Faro, Portugal}

\begin{abstract}
CPT and Lorentz violation in the photon sector is described within
the minimal Standard-Model Extension by a dimension-3 Chern-Simons-like
operator parametrized by a four-vector parameter $k_{AF}$
that has been very tightly bounded by astrophysical observations.
On the other hand, in the context of the $SU(2)\times U(1)$ electroweak gauge sector of the Standard-Model Extension, CPT and Lorentz violation
is described similarly, by dimension-3 operators
parametrized by four-vector parameters $k_1$ and $k_2$.
In this work, we investigate in detail the effects of the resulting
CPT and Lorentz violation in the photon and $Z$-boson sectors upon
electroweak-symmetry breaking.
In particular, we show that for the photon sector the relevant Lorentz-violating
effects are described at lowest order by the $k_{AF}$ term,
but that there are higher-order momentum-dependent effects due to photon-$Z$ boson mixing.
As bounds on CPT and Lorentz violation in the $Z$ sector are relatively weak,
these effects could be important phenomenologically.
We investigate these effects in detail in this work.
\end{abstract}

\maketitle

\section{Introduction}
The main motivation to search for departures of relativity is that various candidate
theories of gravity may allow for spontaneous Lorentz violation
\cite{qgmodels1a,qgmodels1b,qgmodels2,qgmodels3,qgmodels4,qgmodels5}.
A framework that allows incorporation of Lorentz-violating effects into the
Standard Model is the Standard-Model Extension (SME) \cite{sme1,sme2},
an effective field theory containing all Lorentz- and CPT-violating operators that are allowed by the remaining Standard-Model symmetries.
The SME also contains all CPT-violating operators,
since in any local interacting quantum field theory CPT violation implies
Lorentz violation \cite{Gre02}.
While the SME, being an effective field theory, consists of terms of arbitrary
mass dimension, we will consider in this work the superficially renormalizable part
of the SME, called the minimal SME (mSME).
When restricted to the electroweak part of the gauge sector,
CPT and Lorentz violation is parametrized by two four-vector coefficients
$k_1^\mu$ and $k_2^\mu$, corresponding to the $U(1)_Y$ and $SU(2)$ gauge sectors,
respectively.
Each of them multiplies a CPT-violating Chern-Simons-like term
of mass dimension three.
The $k_2$ term acts to modify the kinetic term of the $W$-boson sector.
For massive gauge bosons it has been shown in \cite{ColladayMcDonaldNoordmansPotting}
that even in the presence of such a CPT-violating term covariant quantization
of the gauge field theory can be carried out consistently.
In \cite{Wboson} it was shown that the dispersion relation for $W$ bosons
is modified in such a way as to allow for Cherenkov-like emission by high-energy
fermions.
This effect allows for the derivation of a bound on the $k_2$ parameter
by considering observational data on protons in ultra-high-energy cosmic rays.

For the $Z$ boson and the photon sectors the situation is more complicated,
due to the mixing that is provoked by electroweak symmetry breaking.
As we will show in this paper,
the $Z$ boson and the photon each receive CPT-violating
contributions from both the $k_1$ and $k_2$ terms.
To lowest order, the effect on the photon sector is to provide a Chern-Simons
term with a coefficient $k_{AF}$ that is a certain linear combination of
$k_1$ and $k_2$.
The effect of such a term has been studied long ago in \cite{CarrollFieldJackiw},
where an extremely strong bound on $k_{AF}$ was derived
from astrophysical observations.
(We refer to \cite{datatables} for an up-to-date list of
experimental and observational bounds on SME parameters.)

However, as we will show in this work,
it turns out that there are other,
higher-order effects arising from the $k_1$ and $k_2$ terms
that are provoked by the mixing between the photon and the $Z$-boson sectors.
That is, even when one takes the extremely tight bound on $k_{AF}$ at face value,
the mixing with the $Z$ sector provokes the appearance of other
CPT-even and CPT-odd operators into the photon sector.

In this work we will study in detail how this comes about.
We will first use an intuitive perturbative argument to show how the mixing
process arises at low energy for the photon sector.
Then we consider both the photon and $Z$-boson sectors together,
and derive the polarization vectors and the dispersion relations.
We study the latter in detail, showing how the Lorentz violation can give
rise to negative-energy states.
We also analyze the group velocity, and show it never exceeds $c$,
assuring that causality is guaranteed.
Finally, we discuss an application of the extended hamiltonian formalism.

\section{CPT and Lorentz-violation at low energy}

The Lagrangian for the Standard-Model gauge fields $A$ and $Z$,
including the CPT-odd Lorentz-violating terms of the mSME, is given by
\begin{equation}
\Ll_{AZ} = \frac12 A_\mu D^{\mu\nu} A_\nu + \frac12 Z_\mu D^{\mu\nu} Z_\nu
+ \frac12 m_Z^2 Z_\mu Z^\mu + 
\frac12 \epsilon_{\mu\nu\rho\sigma}\left(
k_{AF}^\mu A^\nu F^{\rho\sigma} + k_{ZZ}^\mu Z^\nu Z^{\rho\sigma} +
2k_{\mbox{\tiny mix}}^\mu Z^\nu F^{\rho\sigma}\right)
\label{AZ-Lagrangian}
\end{equation}
where $D^{\mu\nu} = \partial^2 \eta^{\mu\nu}-(1-\xi^{-1})\partial^\mu \partial^\nu$,
and
\begin{eqnarray}
k_{AF}^\mu &=& 2c_w^2 k_1^\mu + s_w^2 k_2^\mu\,,\\
k_{ZZ}^\mu &=& 2s_w^2 k_1^\mu + c_w^2 k_2^\mu\,,\\
k_{\mbox{\tiny mix}}^\mu &=& c_w s_w(k_2^\mu - 2k_1^\mu)\,,
\end{eqnarray}
with $c_w = \cos \theta_w$ and $s_w = \sin \theta_w$.
Because of the mixing term in (\ref{AZ-Lagrangian}),
the asymptotic fields are no longer $A^\mu$ and $Z^\mu$,
but appropriate linear combinations of these fields and their derivatives.
In the remainder of this paper, we will study the physical consequences of this.

As is well known, the photon parameter $k_{AF}$ is extremely strongly bound
by astrophysical observations,
by observing synchrotron radiation from radio galaxies
(at $\lambda \sim 10~\mbox{cm}$) \cite{CarrollFieldJackiw},
as well as from from CMB polarimetry studies
($\lambda \sim 1~\mbox{mm}$) \cite{KosteleckyMewes1}.
In both cases the listed bounds on the components of $k_{AF}$ are of the order
of $10^{-42}$~GeV.
In view of the above mixing, however,
these bounds should presumably apply to the asymptotic states,
rather than to the photon field $A^\mu$.
That is, one should first diagonalize the kinetic Lagrangian in the full $A$, $Z$
field space and then see which Lorentz-violating parameter applies
to the massless propagating degrees of freedom.

Rather than performing such a full diagonalization,
we will first do a perturbative analysis of the photon propagator.
As a first step, we will only consider the $k_{AF}$ parameter.
Up to third order in $k_{AF}$, the photon propagator can be expanded as 
\begin{equation}
{\begin{picture}(30,15)(0,0)
    \SetWidth{0.7}
    \SetColor{Black}
    \Photon[sep=4,clock](0,4)(30,4){2}{4} 
    \Text(5,-3)[]{$A$} \Text(25,-3)[]{$A$} 
 \end{picture}}
\quad + \quad
 {\begin{picture}(60,15)(0,0)
    \SetWidth{0.7}
    \SetColor{Black}
    \Photon[sep=4,clock](0,4)(60,4){2}{8} 
    \Vertex(30,4){2}
    \Text(30,12)[]{$k_{AF}$}
    \Text(5,-3)[]{$A$} \Text(55,-3)[]{$A$} 
 \end{picture}} 
\quad + \quad
{\begin{picture}(90,15)(0,0)
    \SetWidth{0.7}
    \SetColor{Black}
    \Photon[sep=4,clock](0,4)(90,4){2}{12} 
    \Vertex(30,4){2} \Vertex(60,4){2}
    \Text(30,12)[]{$k_{AF}$} \Text(60,12)[]{$k_{AF}$}
    \Text(5,-3)[]{$A$} \Text(45,-3)[]{$A$} \Text(85,-3)[]{$A$}
 \end{picture}}
\quad + \quad
{\begin{picture}(120,15)(0,0)
    \SetWidth{0.7}
    \SetColor{Black}
    \Photon[sep=4,clock](0,4)(120,4){2}{16} 
    \Vertex(30,4){2} \Vertex(60,4){2} \Vertex(90,4){2}
    \Text(30,12)[]{$k_{AF}$} \Text(60,12)[]{$k_{AF}$} \Text(90,12)[]{$k_{AF}$}
    \Text(5,-3)[]{$A$} \Text(45,-3)[]{$A$} \Text(75,-3)[]{$A$} \Text(115,-3)[]{$A$}
 \end{picture}}
\label{expand_prop_kAF}
\end{equation}
Truncating the external legs, the first-order diagram is represented by
\begin{equation}
\begin{picture}(0,15)(0,0)
    \SetWidth{0.7}
    \SetColor{Black}
    \Vertex(0,4){2}
    \Text(0,12)[]{$k_{AF}$}
 \end{picture}
\quad \to \quad -\frac{i}{2}\epsilon_{\mu\nu\alpha\beta}k_{AF}^\alpha p^\beta \,.
\label{first-order-diagram}
\end{equation}
For the truncated second-order diagram one finds the CPT-even expression
\begin{equation}
{\begin{picture}(30,15)(0,0)
    \SetWidth{0.7}
    \SetColor{Black}
    \Photon[sep=4,clock](0,4)(30,4){2}{4} 
    \Vertex(0,4){2} \Vertex(30,4){2}
    \Text(0,12)[]{$k_{AF}$} \Text(30,12)[]{$k_{AF}$}
    \Text(15,-3)[]{$A$}
 \end{picture}}
 \quad \to \quad -\frac{i}{4}\frac{1}{p^2+i\epsilon}
 \left[k_{AF}^2(p^2\eta_{\mu\nu}-p_\mu p_\nu)
 - k^A_{F,\mu\alpha\nu\beta}p^\alpha p^\beta\right]
\end{equation}
where the Lorentz-violating tensor coefficient
\begin{equation}
k^A_{F,\mu\alpha\nu\beta} =
\eta_{\mu\nu}k_{AF,\alpha}k_{AF,\beta} - \eta_{\alpha\nu}k_{AF,\mu}k_{AF,\beta} -
\eta_{\mu\beta}k_{AF,\alpha}k_{AF,\nu} + \eta_{\alpha\beta}k_{AF,\mu}k_{AF,\nu}
\label{k(A)_F}
\end{equation}
has the symmetries of the Riemann tensor and
can be interpreted as a contribution to the non-birefringent sector of the
$CPT$-even Lorentz-violating tensor coefficient $k_F$.
At third order, we find
\begin{eqnarray}
{\begin{picture}(60,15)(0,0)
    \SetWidth{0.7}
    \SetColor{Black}
    \Photon[sep=4,clock](0,4)(60,4){2}{8} 
    \Vertex(0,4){2} \Vertex(30,4){2} \Vertex(60,4){2}
    \Text(0,12)[]{$k_{AF}$} \Text(30,12)[]{$k_{AF}$} \Text(60,12)[]{$k_{AF}$}
     \Text(15,-3)[]{$A$} \Text(45,-3)[]{$A$}
 \end{picture}}
 \quad &\to& \quad -\frac{i}{8}\frac{1}{(p^2+i\epsilon)^2}\epsilon_{\mu\nu\alpha\beta}k_{AF}^\alpha p^\beta\left(k_{AF}^2p^2-(k_{AF}\cdot p)^2\right)\nonumber\\[8pt]
 &=& \quad \frac14 \frac{k_{AF}^2p^2-(k_{AF}\cdot p)^2}{(p^2+i\epsilon)^2}
 \quad\times \qquad
  {\begin{picture}(0,15)(0,0)
    \SetWidth{0.7}
    \SetColor{Black}
    \Vertex(0,4){2}
    \Text(0,12)[]{$k_{AF}$}
    \,.
 \end{picture}}
\end{eqnarray}
This represents a CPT-violating operator proportional to the first-order term
(\ref{first-order-diagram}), with momentum-dependent proportionality factor.
Altogether, up to third order in $k_{AF}$,
the corrections to the photon propagator yield
\begin{equation}
 -\frac{i}{2}\epsilon_{\mu\nu\alpha\beta}k_{AF}^\alpha p^\beta
 \left(1+\frac14 \frac{k_{AF}^2p^2-(k_{AF}\cdot p)^2}{(p^2+i\epsilon)^2}\right)
-\frac{i}{4}\frac{1}{p^2+i\epsilon}
 \left(k_{AF}^2(p^2\eta_{\mu\nu}-p_\mu p_\nu)
 - k^A_{F,\mu\alpha\nu\beta}p^\alpha p^\beta\right)\,.
\label{kAF_3rd_order}
\end{equation}
There is one problematic aspect one notices in the expansion (\ref{kAF_3rd_order}):
starting at quadratic order, the corrections diverge as $p^2\to 0$.
This naturally happens because $p^2=0$ amounts to the unperturbed photon mass-shell
condition.
For nonzero $k_{AF}$ the mass-shell condition will be modified, but nevertheless,
for very small $k_{AF}$, $p^2$ will be very close to zero on the mass shell.
This means that it is unclear whether the successive terms in the expansion
(\ref{kAF_3rd_order}) converge when the photon is taken on the mass shell.

For this reason, a better method is to simply work with the full
$k_{AF}$-modified photon propagator,
for which an explicit and well-defined expression
has been derived in \cite{ColladayMcDonaldNoordmansPotting} (see eq.~(46)).
It can be checked easily that an order-by-order expansion of
expression (46) of \cite{ColladayMcDonaldNoordmansPotting} yields,
up to third order, the expression (\ref{kAF_3rd_order}).

Having incorporated fully the effect of the $k_{AF}$ term,
we will now go ahead and include
the last two CPT-odd terms of the Lagrangian (\ref{AZ-Lagrangian})
parametrized by the Lorentz-violating parameters $k_{ZZ}$ and $k_{\mbox{\tiny mix}}$.
Representing the $k_{AF}$-modified photon propagator by a double wavy line,
the full modified photon propagator is given by the expansion
\begin{equation}
{\begin{picture}(30,15)(0,0)
    \SetWidth{0.7}
    \SetColor{Black}
    \Photon[double,sep=4,clock](0,4)(30,4){2}{4} 
 \end{picture}}
\quad + \quad
 {\begin{picture}(60,15)(0,0)
    \SetWidth{0.7}
    \SetColor{Black}
    \Photon[double,sep=4,clock](0,4)(60,4){2}{8} 
    \CCirc(30,4){5}{Black}{Gray}
 \end{picture}} 
\quad + \quad
{\begin{picture}(90,15)(0,0)
    \SetWidth{0.7}
    \SetColor{Black}
    \Photon[double,sep=4,clock](0,4)(90,4){2}{12} 
    \CCirc(30,4){5}{Black}{Gray}    \CCirc(60,4){5}{Black}{Gray}
 \end{picture}}
\quad + \quad
{\begin{picture}(120,15)(0,0)
    \SetWidth{0.7}
    \SetColor{Black}
    \Photon[double,sep=4,clock](0,4)(120,4){2}{16} 
    \CCirc(30,4){5}{Black}{Gray} \CCirc(60,4){5}{Black}{Gray}
        \CCirc(90,4){5}{Black}{Gray}
 \end{picture}}
\end{equation}
where the gray blob stands for the mixing terms
\begin{equation}
 {\begin{picture}(0,15)(0,0)
    \SetWidth{0.7}
    \SetColor{Black}
    \CCirc(0,4){5}{Black}{Gray}
 \end{picture}} 
\qquad = \qquad
{\begin{picture}(30,15)(0,0)
    \SetWidth{0.7}
    \SetColor{Black}
    \Photon[sep=4,clock](0,4)(30,4){2}{4} 
    \Vertex(0,4){2} \Vertex(30,4){2}
    \Text(0,12)[]{$k_{\mbox{\tiny mix}}$} \Text(30,12)[]{$k_{\mbox{\tiny mix}}$}
    \Text(15,-3)[]{$Z$}
 \end{picture}}
 \quad + \quad
 {\begin{picture}(60,15)(0,0)
    \SetWidth{0.7}
    \SetColor{Black}
    \Photon[sep=4,clock](0,4)(60,4){2}{8} 
    \Vertex(0,4){2} \Vertex(30,4){2} \Vertex(60,4){2}
    \Text(0,12)[]{$k_{\mbox{\tiny mix}}$} \Text(30,12)[]{$k_{ZZ}$} \Text(60,12)[]{$k_{\mbox{\tiny mix}}$}
    \Text(15,-3)[]{$Z$} \Text(45,-3)[]{$Z$}
 \end{picture}}
 \quad + \quad
 {\begin{picture}(90,15)(0,0)
    \SetWidth{0.7}
    \SetColor{Black}
    \Photon[sep=4,clock](0,4)(90,4){2}{12} 
    \Vertex(0,4){2} \Vertex(30,4){2} \Vertex(60,4){2} \Vertex(90,4){2}
    \Text(0,12)[]{$k_{\mbox{\tiny mix}}$} \Text(30,12)[]{$k_{ZZ}$} \Text(60,12)[]{$k_{ZZ}$} \Text(90,12)[]{$k_{\mbox{\tiny mix}}$}
    \Text(15,-3)[]{$Z$} \Text(45,-3)[]{$Z$} \Text(75,-3)[]{$Z$}
 \end{picture}}
     \quad + \quad \ldots  
\label{grayBlob} 
\end{equation}
The first term, quadratic in $k_{\mbox{\tiny mix}}$,
represents the operator
\begin{equation}
{\begin{picture}(30,15)(0,0)
    \SetWidth{0.7}
    \SetColor{Black}
    \Photon[sep=4,clock](0,4)(30,4){2}{4} 
    \Vertex(0,4){2} \Vertex(30,4){2}
    \Text(0,12)[]{$k_{\mbox{\tiny mix}}$} \Text(30,12)[]{$k_{\mbox{\tiny mix}}$}
    \Text(15,-3)[]{$Z$}
 \end{picture}}
 \quad \to \quad -\frac{i}{4}\frac{1}{p^2-m_Z^2+i\epsilon}
 \left[k_{\mbox{\tiny mix}}^2(p^2\eta_{\mu\nu}-p_\mu p_\nu)
 - k^{\mbox{\tiny mix}}_{F,\mu\alpha\nu\beta}p^\alpha p^\beta\right]
\end{equation}
with 
\begin{equation}
k^{\mbox{\tiny mix}}_{F,\mu\alpha\nu\beta} =
\eta_{\mu\nu}k_{\mbox{\tiny mix},\alpha}k_{\mbox{\tiny mix},\beta} - \eta_{\alpha\nu}k_{\mbox{\tiny mix},\mu}k_{\mbox{\tiny mix},\beta} -
\eta_{\mu\beta}k_{\mbox{\tiny mix},\alpha}k_{\mbox{\tiny mix},\nu} + \eta_{\alpha\beta}k_{\mbox{\tiny mix},\mu}k_{\mbox{\tiny mix},\nu} \,.
\end{equation}
This operator contributes to the non-birefringent sector of the $CPT$-even
Lorentz-violating tensor coefficient $k_F$, analogously to (\ref{k(A)_F}).
The third-order term
\begin{equation}
{\begin{picture}(60,15)(0,0)
    \SetWidth{0.7}
    \SetColor{Black}
    \Photon[sep=4,clock](0,4)(60,4){2}{8} 
    \Vertex(0,4){2} \Vertex(30,4){2} \Vertex(60,4){2}
    \Text(0,12)[]{$k_{\mbox{\tiny mix}}$} \Text(30,12)[]{$k_{ZZ}$} \Text(60,12)[]{$k_{\mbox{\tiny mix}}$}
    \Text(15,-3)[]{$Z$} \Text(45,-3)[]{$Z$}
 \end{picture}}
 \quad\to\quad
 \frac{-i}{8}\frac{(k_{\mbox{\tiny mix}}\cdot k_{ZZ})p^2 -
 (k_{\mbox{\tiny mix}}\cdot p)(k_{ZZ}\cdot p)}{(p^2-m_Z^2+i\epsilon)^2}
 \epsilon_{\alpha\beta\mu\nu}k_{\mbox{\tiny mix}}^\alpha p^\beta 
 \label{mix-Z-mix}
\end{equation}
represents the lowest-order CPT-odd contribution to the full photon
propagator due to the mixing with the $Z$ boson.
The effect of the operator (\ref{mix-Z-mix}) on the photon propagator
is to modify the original $k_{AF}$ term by changing the vector
\begin{equation}
k_{AF}^\mu \to \tilde k_{AF}^\mu
\end{equation}
with
\begin{equation}
\tilde k_{AF}^\mu = k_{AF}^\mu + k_{\mbox{\tiny mix}}^\mu \,
\frac{(k_{\mbox{\tiny mix}}\cdot k_{ZZ})p^2-(k_{\mbox{\tiny mix}}\cdot p)(k_{ZZ}\cdot p)}{4(p^2-M_Z^2)^2}
\approx
k_{AF}^\mu - k_{\mbox{\tiny mix}}^\mu \,
\frac{(k_{\mbox{\tiny mix}}\cdot p)(k_{ZZ}\cdot p)}{4M_Z^4}
\label{eff-kAF}
\end{equation}
where in the last equation we used $p^2\approx 0$ as a very good approximation
for the photon dispersion relation.

Clearly, the strong observational bounds referred to above apply to the components
of $\tilde k_{AF}$, not to $k_{AF}$.
Thus, it is in principle conceivable that the bound does not apply to
(or is much weaker for) $k_{AF}$, if somehow the effects of the two terms
in (\ref{eff-kAF}) in the relevant CPT-violating processes involved in the
astrophysical observations were to (partially) cancel.
In that case it might be conceivable that a strong bound on $\tilde k_{AF}$
might not imply an equally strong bound on $k_{AF}$.
Note, however, that the observational bounds on
$\tilde k_{AF}$ have been established for at least two very different momentum
scales, namely at radio frequencies as well as for CMB microwave frequencies.
As the two terms in (\ref{eff-kAF}) have very different momentum structures,
it is not possible that a fortuitous cancelation of their effects would
occur for both sets of momenta. 
Thus we conclude that the strong observational bounds apply for $k_{AF}$
as well as for the second term in (\ref{eff-kAF}).

The higher-order terms in Eq.~\eqref{grayBlob} can only be ignored at energies where $k_X\cdot p$ can be considered small (with $X \in \{ZZ, \mathrm{mix}\}$). This will be the case for most applications, notably at the energies that are used to determine bounds on $k_{AF}$. However, when one wants to consider Cherenkov-like processes $k_X\cdot p$ will have to be of order $m_Z^2$ and the higher-order terms cannot be neglected. For such high-energy processes, we cannot take the perturbative approach.

Since the bounds on $k_{AF}$ are obtained at comparatively low energies, we can use the fact that these bounds are so stringent to neglect $k_{AF}$ with respect to $k_{ZZ}$.
Therefore, from now on, we will consider the case where $k^\mu_{AF} = 0$,
such that $k^\mu_{\rm mix} = \frac{1}{2}\tan(2\theta_w)k^\mu_{ZZ}$.
In that case, we can write Lagrangian $\Ll_{AZ}$ in Eq.~\eqref{AZ-Lagrangian}
in terms of an eight-component ``bi-four-vector'' $\Aa$, given by
\begin{equation}
\Aa^\mu  =  \left(\begin{array}{c} A^\mu \\ Z^\mu \end{array}\right)\ . \\
\label{eightvector}
\end{equation}
The Lagrangian becomes:
\begin{equation}
\Ll_{AZ} = \frac{1}{2}\Aa^T_\mu D^{\mu\nu}\Aa_\nu + \frac{1}{2}\Aa^T_\mu \Mm\Aa^\mu - \Aa^T_\mu \Xx k^{\mu\nu}_{ZZ} \Aa_\nu\ ,
\label{L_AZ}
\end{equation}
where
$D^{\mu\nu} = \partial^2 \eta^{\mu\nu} - (1-\xi^{-1})\partial^\mu \partial^\nu$,
$k_{ZZ}^{\mu\nu} = \epsilon^{\mu\nu\rho\sigma}(k_{ZZ})_\rho\partial_\sigma$,
and
\begin{subequations}
\begin{eqnarray}
\Mm & = & \left(\begin{array}{cc}0 & 0 \\ 0 & m_Z^2\end{array}\right)\ , \\
\Xx & = & \left(\begin{array}{cc} 0 & \tfrac{1}{2}\tan(2\theta_w) \\ \tfrac{1}{2}\tan(2\theta_w) & 1 \end{array}\right)\ .
\end{eqnarray}
\end{subequations}

\section{Polarization vectors}

Before treating the solutions of the equation of motion corresponding to Eq.\  \eqref{L_AZ},
let us review the simpler case of a single four-vector:
\begin{equation}
\Ll_{Z} = \frac{1}{2}Z_\mu D^{\mu\nu}Z_\nu + \frac{1}{2}Z_\mu \Mm Z^\mu 
- Z_\mu k^{\mu\nu}_{ZZ} Z_\nu
\label{L_Z}
\end{equation}
that was treated in detail in \cite{ColladayMcDonaldNoordmansPotting}.
In momentum space, the equation of motion of Eq.\ \eqref{L_Z} reads
\begin{equation}
\left[(p^2-m_Z^2)\eta^\mu{}_\nu-(1-\xi^{-1})p^\mu p_\nu
-2i\epsilon^{\alpha\beta\mu}{}_\nu(k_{ZZ})_\alpha p_\beta\right]
e^{(\lambda)\nu}(\vec p) \equiv S^\mu{}_\nu e^{(\lambda)\nu}(\vec p) = 0\ ,
\label{eom-Z}
\end{equation}
where $e^{(\lambda)\nu}(\vec p)$ are the eigenvectors of the
equation-of-motion operator $S^\mu{}_\nu$.
The index $\lambda$ runs over $0,3,+,-$,
labeling the gauge mode, and three physical modes, respectively.
It can be shown \cite{ColladayMcDonaldNoordmansPotting} that
the eigenvalues of $S^\mu{}_\nu$ are given by the expressions
\begin{subequations}
\begin{eqnarray}
\Lambda_0(p) &=& \frac{1}{\xi}(p^2-\xi m_Z^2)\ , \\
\Lambda_3(p) &=& p^2 - m_Z^2\ , \\
\Lambda_\pm(p) &=& p^2 - m_Z^2 \pm 2\sqrt{(p\cdot k_{ZZ})^2 - p^2 k_{ZZ}^2}\ .
\end{eqnarray}
\label{eigenvalues-Z}
\end{subequations}
These observer-scalar functions of $p^\mu$ and $k_{ZZ}^\mu$ define the dispersion relations for each of the polarization modes by fixing $\Lambda_\lambda(p) = 0$. 
The corresponding eigenvectors $e^{(\lambda)\nu}(\vec p)$ can be constructed
explicitly \cite{ColladayMcDonaldNoordmansPotting}.

Let us now consider the equation of motion of Lagrangian $\Ll_{AZ}$
in momentum space, which can be written as:
\begin{equation}
\left[(p^2-\Mm)\eta^\mu{}_\nu-(1-\xi^{-1})p^\mu p_\nu
-2i\Xx\epsilon^{\alpha\beta\mu}{}_\nu(k_{ZZ})_\alpha p_\beta\right]
\tau^{(\lambda)\nu}_\sigma  \equiv \Ss^\mu{}_\nu \tau^{(\lambda)\nu}_\sigma  = 0
\end{equation}
where $\tau^{(\lambda)\nu}_\sigma(\vec p)$,
with $\sigma = \pm 1$ and $\lambda \in \{0,3,+,-\}$,
are the eigenvectors of the equation-of-motion operator $\Ss^\mu{}_\nu$.
We will make the following ansatz for the eight-component polarization vectors:
\begin{equation}
\tau^{(\lambda)\mu}_\sigma = \left(\begin{array}{c} \alpha_\sigma^\lambda e^{(\lambda)\mu} \\ \beta_\sigma^{\lambda} e^{(\lambda)\mu}\end{array}\right)\ .
\end{equation}
Here $\alpha_\sigma^\lambda$ and $\beta_\sigma^{\lambda}$ are scalars
that need to be determined.
The four-vectors $e^{(\lambda)\mu}$ are the eigenvectors of the equation-of-motion
operator $S^\mu{}_\nu$ introduced in Eq.\ \eqref{eom-Z}.
The fact that this ansatz works, hinges on the fact that there is
only one Lorentz-violating four-vector in Eq.\ \eqref{eom-Z},
i.e., on the fact that we put $k^\mu_{AF}$ to zero.
Allowing for the possibility that $k^\mu_{AF} \ne 0$
would introduce a second independent Lorentz-violating four-vector,
making the technical analysis much more complicated.

It turns out that the eigenvectors $\tau_\sigma^{(\lambda)\mu}$
correspond to eigenvalues $\Omega_{\sigma}^\lambda(p)$ given by
\begin{subequations}
\begin{eqnarray}
\Omega_{+1}^0(p) &=& \frac{1}{\xi}p^2 \ ,\\
\Omega_{-1}^0(p) &=& \frac{1}{\xi}(p^2 - \xi m_Z^2) \ ,\\
\Omega_{+1}^3(p) &=& p^2 \ ,\label{disprelation_long_photon}\\
\Omega_{-1}^3(p) &=& p^2 - m_Z^2 \ ,\\
\Omega_{+1}^\pm(p) &=& p^2 - \frac{1}{2}m_Z^2 \pm \delta(p)+ \frac{1}{2}\sqrt{(m_Z^2\mp 2\delta(p))^2+ 4\tan^2(2\theta_w)\delta(p)^2}\ ,
\label{disp_transverse1}\\
\Omega_{-1}^\pm(p) &=& p^2 - \frac{1}{2}m_Z^2 \pm \delta(p)- \frac{1}{2}\sqrt{(m_Z^2\mp 2\delta(p))^2+ 4\tan^2(2\theta_w)\delta(p)^2}\ .
\label{disp_transverse2}
\end{eqnarray}
\label{disprelations}%
\end{subequations}
with $\delta(p) = \sqrt{(p\cdot k_{ZZ})^2 - p^2k_{ZZ}^2} = \frac{2}{\tan(2\theta_w)}\sqrt{(p\cdot k_{\rm mix})^2 - p^2k_{\rm mix}^2}$. For ``small'' energies we can expand the square roots of the final two expressions and obtain
\begin{subequations}
\begin{eqnarray}
\Omega_{+1}^\pm(p) &=& p^2 +\tan^2(2\theta_w)\frac{\delta(p)^2}{m_Z^2} \pm 2\tan^2(2\theta_w)\frac{\delta(p)^3}{m_Z^4} + \cdots \ ,\\
\Omega_{-1}^\pm(p) &=& p^2 -m_Z^2  \pm 2\delta(p) - \tan^2(2\theta_w)\frac{\delta(p)^2}{m_Z^2} \mp 2\tan^2(2\theta_w)\frac{\delta(p)^3}{m_Z^4} + \cdots \ ,
\end{eqnarray}
\label{expansions}
\end{subequations}
From Eqs.~\eqref{disprelations} and \eqref{expansions} it is clear that the
$\sigma=+1$ modes are massless and the $\sigma=-1$ modes are massive,
at least in the limit of small Lorentz violation and low energies.
We will therefore call the former mode the photon and the latter mode the $Z$ boson.
The massless eigenvector corresponding to Eq.\ \eqref{disprelation_long_photon}
corresponds to the longitudinal photon mode.
We expect it to decouple upon applying a BRST quantization procedure,
analogous to the one presented in \cite{ColladayMcDonaldNoordmansPotting}.

The explicit expressions for $\alpha_\sigma^\lambda$ and $\beta_\sigma^\lambda$ are:
\begin{subequations}
\begin{eqnarray}
\alpha_{+1}^{0} & = & \alpha_{+1}^3 = \beta_{-1}^{0} = \beta_{-1}^3 = 1\ , \\
\alpha_{-1}^{0} & = & \alpha_{-1}^3 = \beta_{+1}^{0} = \beta_{+1}^3 = 0\ , \\
\alpha_{\sigma}^{\pm} &=& \frac{\tan(2\theta_w)\delta(p)}{\sqrt{(p^2-\Omega_\sigma^{\pm}(p))^2 + \tan^2(2\theta_w)\delta(p)^2}}\ , \\
\beta_{\sigma}^{\pm} &=& \frac{p^2 - \Omega_\sigma^\pm(p)}{\sqrt{(p^2-\Omega_\sigma^{\pm}(p))^2 + \tan^2(2\theta_w)\delta(p)^2}}\ .
\end{eqnarray}
\end{subequations}
such that $(\alpha_\sigma^\lambda(p))^2 + (\beta_\sigma^\lambda(p))^2 = 1$ and the eigenvectors $\tau_\sigma^{(\lambda)\mu}$, when evaluated at the same four-momentum, obey
\begin{equation}
\tau_\sigma^{(\lambda)*}(p)\cdot \tau_{\sigma'}^{(\lambda')}(p) = \delta_{\sigma\sigma'}g^{\lambda\lambda'}\ ,
\end{equation}
where $g^{\lambda\lambda'} = e^{(\lambda)*}\cdot e^{(\lambda')}$. 
A low-energy approximation for $\alpha_\sigma^\pm$ and $\beta_\sigma^\pm$ reads
\begin{subequations}
\begin{eqnarray}
\alpha_{+1}^{\pm} &\approx & \beta_{-1}^{\pm} \approx 1 -\frac{\tan^2(2\theta_w)\delta(p)^2}{2m_Z^2} \mp \frac{2\tan^2(2\theta_w)\delta(p)^3}{m_Z^6}+ \cdots\ , \\
\alpha_{-1}^{\pm} &\approx & - \beta_{+1}^{\pm} \approx \tan(2\theta_w)\left(\frac{\delta(p)}{m_Z^2} \pm \frac{\delta(p)^2}{m_Z^4} + \frac{(8 - 3\tan^2(2\theta_w))\delta(p)^3}{2m_Z^6} + \cdots\right) \ ,
\end{eqnarray}
\end{subequations}
where the dots stand for terms of at least order $\mathcal{O}(\delta(p)^2)$.
Comparing to Eq.~\eqref{eightvector},
we see that in the limit of small Lorentz violation and low energy the $\sigma=+1$ ($\sigma = -1)$ mode corresponds to the conventional photon ($Z$ boson).
In the mentioned limits the modes thus also couple correctly to the conventional fermion currents.

\section{Analysis of the dispersion relation}

In this section, we analyse the dispersion relations for the different particle modes.
In particular, we address the question if the Lorentz-violating dispersion relations $\Omega_\sigma^\pm(p)=0$ have two roots for each mode.
Moreover, we want to determine if any of the roots is degenerate.

\begin{figure}[h]
	\centering
	\includegraphics[width=\textwidth]{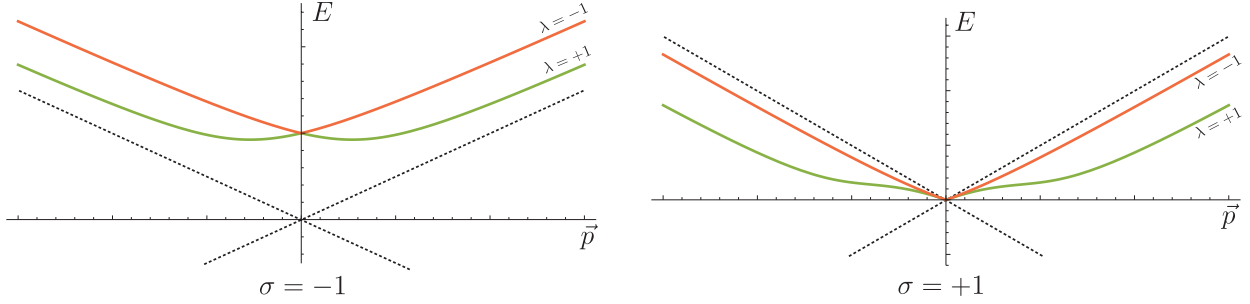}
	\caption{Plots of sample solutions to the $\lambda=\pm$ dispersion relations for the $Z$ boson (left) and photon (right). The size of the Lorentz-violating effects has been exaggerated for illustrational purposes.}
	\label{fig:disrel3plot}
\end{figure}

The relevant Lorentz-violating dispersion relations can be written as
\begin{equation}
\Omega_\sigma^\lambda(p) =  p^2 - \frac{1}{2}m_Z^2 + \lambda \delta(p)+ \frac{\sigma}{2}\sqrt{(m_Z^2 - 2\lambda \delta(p))^2+ 4\tan^2(2\theta_w)\delta(p)^2} = 0\ ,
\label{disprel1}
\end{equation}
with $\lambda\in \{+,-\}$ and $\sigma \in \{-1,+1\}$. First of all, we notice that $\Omega_\sigma^\lambda(p) = \Omega_\sigma^\lambda(-p)$. Therefore, if the dispersion relation has two solutions (as we will show below), the usual redefinition of the negative-energy states will map them onto each other, i.e. particles and antiparticles have the same energy.
Furthermore, since the square root (without $\sigma$) on the right-hand side is always larger than or equal to $|\tfrac{1}{2}m_Z^2 - \lambda\delta(p)|$, the sign of $p^2$ is determined by $\sigma$, i.e.
\begin{equation}
p^2 \left\{\begin{array}{ccc} \leq 0 & {\rm for} & \sigma = +1 \\
> 0 & {\rm for} & \sigma = -1
\end{array}\right.\ .
\label{spacetimelike}
\end{equation}
This shows that the photon mode always has spacelike (or lightlike) momenta, while the $Z$-boson momentum is always timelike (see Fig.~\ref{fig:disrel3plot}). This is to be contrasted with what happens when one does not consider the Lorentz-violating photon--$Z$-boson mixing term, in which case both spacelike and timelike momenta are possible for the photon and the $Z$-boson, as demonstrated in Ref.~\cite{ColladayMcDonaldNoordmansPotting}.
Notice that the equality in Eq.~\eqref{spacetimelike} (for $\sigma = +1$) can only hold if $(p\cdot k_{ZZ}) = 0$. This is only possible if $k^\mu_{ZZ}$ is spacelike/lightlike, or if $p = (0,\vec{0})$. Incidentally, $p = (0, \vec{0})$ is always a solution of $\Omega_{+1}^\lambda(p) = 0$, such that the photon dispersion relation always passes through this point in momentum space. Finally, it follows from Eq.~\eqref{spacetimelike} that the roots of $\Omega_{+1}^\lambda(p)$ are different from those of $\Omega_{-1}^\lambda(p)$ (for the same $\vec{p}$), which answers part of the question about the degeneracy of the roots.

\begin{figure}[ht]
	\centering
	\includegraphics[width=\textwidth]{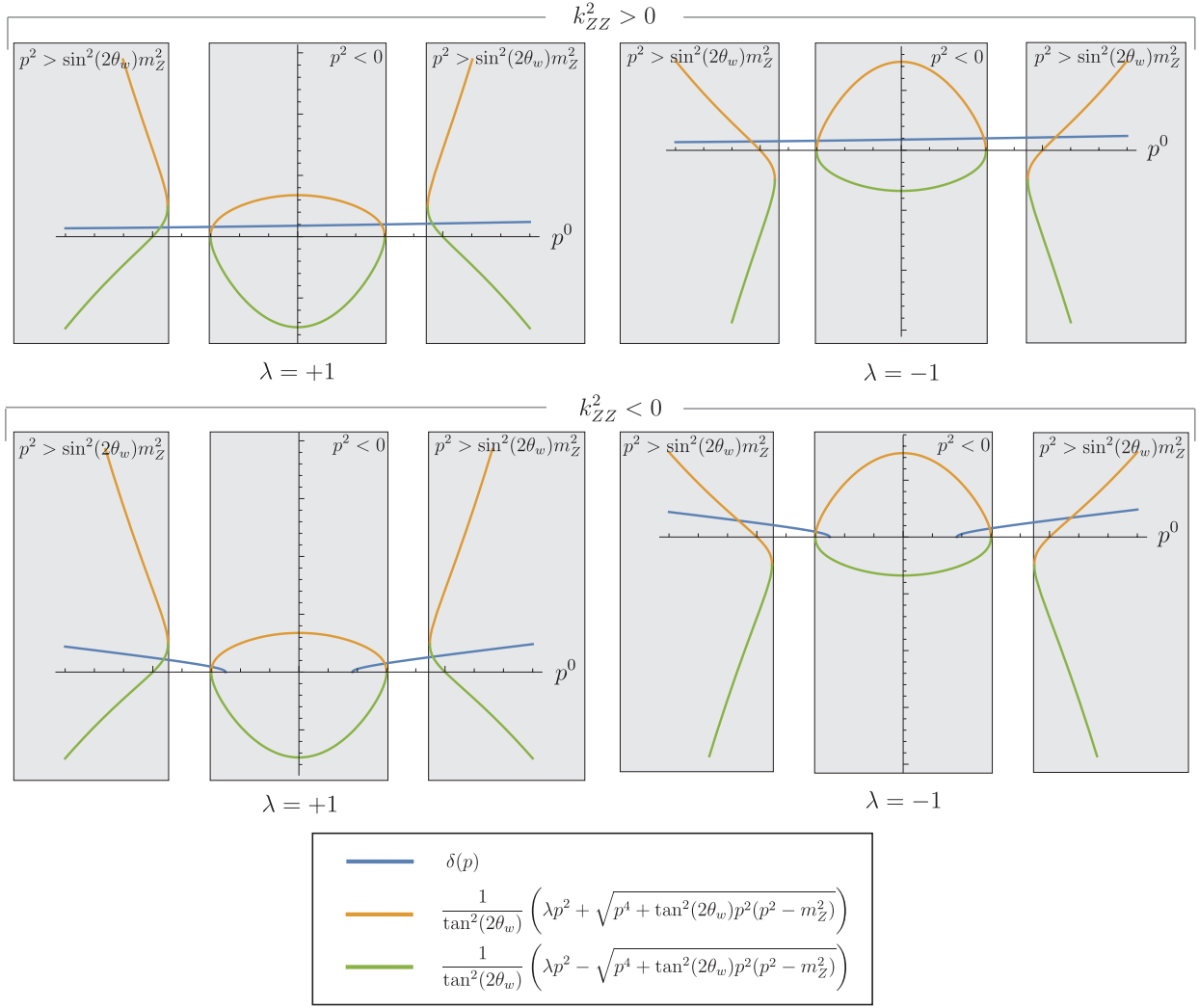}
	\caption{Plots of the left- and right-hand side in Eq.~\eqref{disprel2}.}
	\label{fig:disrel2plot}
\end{figure}

To further address the existence and degeneracy of the solutions of the dispersion relations, we solve Eq.~\eqref{disprel1} for $\delta(p)$ and obtain that to satisfy the dispersion relation, it must hold that
\begin{equation}
\delta(p) = f^\lambda_\pm(p)\ ,
\label{disprel2}
\end{equation}
with
\begin{equation}
f^\lambda_\pm(p) = \frac{1}{\tan^2(2\theta_w)}\left(\lambda p^2 \pm \sqrt{p^4 + \tan^2(2\theta_w)p^2(p^2 - m_Z^2)}\right)\ ,
\end{equation}
where the $\pm$ sign {\it a priori} has nothing to do with either $\sigma$ or $\lambda$. Notice that Eq.~\eqref{disprel2} does not depend on $\sigma$, however, using the constraint in Eq.~\eqref{spacetimelike} we can still select the appropriate roots, corresponding to $\sigma = + 1$ or to $\sigma = -1$.
To clarify this, we have plotted $\delta(p)$ and $f^\lambda_\pm(p)$ as a function of $p^0$ in
Fig.~\ref{fig:disrel2plot},
for spacelike as well as timelike $k_{ZZ}^\mu$ and for $\lambda = +1$ and $\lambda = -1$. It becomes clear that the intersections of the blue line (corresponding to $\delta(p)$) with the yellow line (corresponding to $f^\lambda_\pm(p)$) that fall in the center gray patch (for which $p^2 <0$), correspond to solutions of the dispersion relation with $\sigma = +1$, while the intersections in the leftmost and rightmost gray patches correspond to solutions of the dispersion relations with $\sigma = -1$.

To find the conditions under which the dispersion relations always have two real solutions each, we notice that the derivative of $\delta(p)$ with respect to $p^0$ goes to $|\vec{k}_{ZZ}|$ for $p^0\rightarrow \pm \infty$, while the derivative of $f_\pm(p)$ goes to $\pm \infty$ in that same limit (the sign in $\pm \infty$ corresponds to the sign in $p^0 \rightarrow \pm \infty$, and not to the subscript of $f_\pm(p)$). Furthermore, for spacelike $k_{ZZ}^\mu$, the blue lines end on the horizontal $p^0$ axis where $\delta(p)$ becomes imaginary. This means that, for spacelike $k^\mu_{ZZ}$, the blue lines in the figures will always intersect both of the outer branches of the yellow or the green line in Fig.~\ref{fig:disrel2plot}, unless the starting points of the blue branches lie in the left- and rightmost gray patches. However, it is easy to verify that $\delta(p)$ can only be imaginary if $p^\mu$ is spacelike, showing that the starting points will, in fact, always lie in the middle gray patch. This then shows that --- at least for spacelike $k^\mu_{ZZ}$ --- the blue line intersects the yellow and green lines exactly four times for $\lambda = +1$ and four times for $\lambda = -1$. These intersections correspond to the eight real roots of $\Omega_\sigma^\lambda(p)$ for $\sigma,\lambda \in \{+1, -1\}$ for the case of spacelike $k^\mu_{ZZ}$. 

For timelike $k^\mu_{ZZ}$, the situation is a little different. From the two top plots in Fig.~\ref{fig:disrel2plot} it is clear that the $\sigma = -1$ dispersion relations always have two roots. They correspond to the intersections in the leftmost and rightmost gray patches in the figures. That these intersections always exist follows from considering the derivatives of $\delta(p)$ and $f_\pm(p)$, as we did in the previous paragraph. The intersections in the middle gray patch ($p^2 < 0$), however, are not guaranteed to exist. It could happen that the blue line does not intersect the yellow line-piece in the middle patch if it lies entirely above this line-piece. In that case, the function $\Omega^{\lambda}_{+1}(p)$ has complex roots. There exists, however, an observer-invariant condition to prevent the latter from happening. Following exactly the same procedure as we did in the appendix of Ref.~\cite{ColladayMcDonaldNoordmansPotting},
one finds that there are always two non-degenerate, real roots,
as long as
\begin{equation}
k_{ZZ}^2 < \breve{m}_Z^2\ ,
\label{conditionrealroots}
\end{equation}
with $\breve{m}_Z^2 = \frac{m_Z^2}{1 + \tan^2(2\theta_w)}$.
In that case we can always find a point where the blue line goes below the yellow line in the middle gray patch of Fig.~\ref{fig:disrel2plot}, resulting in two intersections. This shows that the dispersion relations have two solutions each, also for timelike $k^\mu_{ZZ}$.

It is interesting to note that it might happen that the two intersections of the blue and yellow line lie on the same side of the $p^0 = 0$ line. In that case the blue line lies above the yellow one at $p^0 = 0$ and the two roots have the same sign. If they both are on the left ($p^0 < 0$) side, the energy will be negative even after the usual redefinition of the negative-energy states, since one negative state will map onto another negative state in this case. We thus investigate under what conditions the blue line lies above the yellow line at $p^0 = 0$.

At $p^0 = 0$, we get that
\begin{eqnarray}
f^\lambda_+(p^0 = 0) &=& \frac{1}{\tan^2(2\theta_w)}\left(-\lambda\vec{p}^2 + |\vec{p}|\sqrt{\vec{p}^2(1+\tan^ 2(2\theta_w)) + \tan^2(2\theta_w)m_Z^ 2}\right)\ ,\\
\delta(p^0=0) &=& |\vec{p}|\sqrt{k_{ZZ}^2 + \vec{k}^2_{ZZ}\cos^2\theta}\ ,
\end{eqnarray}
where $\theta$ is the angle between $\vec{p}$ and $\vec{k}_{ZZ}$.
Clearly, 
$f_+(p^0 = 0) <  \delta(p^0=0)$ is the situation we are looking for. We thus investigate $f^\lambda_+(p^0 = 0) -  \delta(p^0=0)$ and see when this is negative. This function vanishes for $|\vec{p}| = 0$ and for
\begin{equation}
|\vec{p}|^\lambda_\pm = \lambda \sqrt{k_{ZZ}^2 + \vec{k}_{ZZ}^2\cos^2\theta } \pm \sqrt{(k_{ZZ}^2+\vec{k}_{ZZ}^ 2\cos^2\theta)(1+\tan^2(2\theta_w)) - m_Z^2}\ ,
\label{pinterval}
\end{equation}
(again, the $\pm$ sign is not related to the value of $\sigma$ or $\lambda$). Moreover, $f^\lambda_+(p^0 = 0) -  \delta(p^0=0)$ goes to $\mp \infty$ for $|\vec{p}| \rightarrow \pm \infty$. Depending on the relative size of the two square roots in Eq.~\eqref{pinterval}, these thus define a $|\vec{p}|$-interval for which both roots of the dispersion relation have the same sign. The second square root is always imaginary if
\begin{equation}
(k^ 0_{ZZ})^2 < \breve{m}_Z^2\ .
\label{conditiondifferentsignroots}
\end{equation} 
%
So if this inequality holds, there is no physical $|\vec{p}|$-interval for which $f^\lambda_+(p^0 = 0) <  \delta(p^0=0)$, i.e. the roots are always real and have different signs. After redefining the roots, all energies are positive.

\begin{figure}[ht]
	\centering
	\includegraphics[height=0.65\textheight]{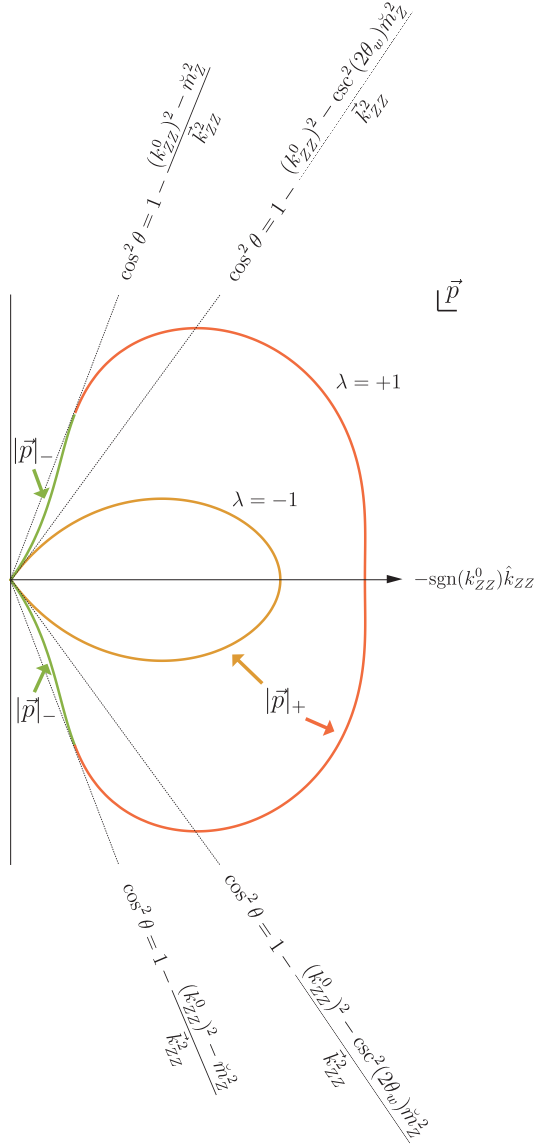}
	\caption{Plots of the photon-momentum region with negative energy.}
	\label{fig:negenergy}
\end{figure}

On the other hand, if $(k^0_{ZZ})^2 > \breve{m}_Z^2$, then negative energies can occur. 
Provided this condition holds,
the second square root in Eq.~\eqref{pinterval} is real if
\begin{equation}
\cos^2\theta > 1 - \frac{(k^0_{ZZ})^2 - \breve{m}_Z^2}{\vec{k}_{ZZ}^2}\ ,
\label{cone1}
\end{equation}
where the second term on the right-hand side is always smaller than one, due to Eq.~\eqref{conditionrealroots}.
The second square root in Eq.~\eqref{pinterval}
is larger than the first 
square root if
\begin{equation}
\cos^2\theta > 1 - \frac{(k^0_{ZZ})^2 - \csc^2(2\theta_w) \breve{m}_Z^2}{\vec{k}_{ZZ}^2}\ .
\label{cone2}
\end{equation}
Furthermore, $k^0_{ZZ}(\vec{p}\cdot \vec{k}_{ZZ}) <0$ is a condition for the energy to be negative after redefinition. This can be seen by realizing that the minimum of $\delta(p)$ lies at $p^0 = k^0_{ZZ}(\vec{p}\cdot \vec{k}_{ZZ}) / \vec{k}_{ZZ}^2$. This thus determines if the minimum, and therefore also the two roots (if they have the same sign), lie to the left or to the right of the $p^0 = 0$ line.
Together with this condition, Eqs.~\eqref{cone1} and \eqref{cone2} both define a momentum cone around $-{\rm sgn}(k_{ZZ}^0)\vec{k}_{ZZ}$. The cone defined by Eq.~\eqref{cone2} is smaller than the one defined by Eq.~\eqref{cone1}. For the $\lambda = -1$ mode, the direction of the photon momentum has to lie within the smaller cone, for the energy to be negative. The absolute momentum then has to obey $0 < |\vec{p}| < |\vec{p}|^{\lambda=-1}_+$. This is depicted in Fig.~\ref{fig:negenergy}.
For the $\lambda = +1$ mode the energy is negative for the same directions if $0 < |\vec{p}| < |\vec{p}|^{\lambda=+1}_+$. However, for $\lambda = +1$ there is an extra set of angles for which the energy can be negative, outside the smaller cone,
but inside the larger cone (see Fig.~\ref{fig:negenergy}).
For these directions the energy is negative if $|\vec{p}|_-^{\lambda=+1} < |\vec{p}| < |\vec{p}|_+^{\lambda = +1}$. Notice that this also means that there are observer frames, where only the $\lambda = +1$ mode can have negative energies. In those frames only the larger cone, defined by Eq.~\eqref{cone1}, exists.

We thus found that each of the four dispersion relations $\Omega_\sigma^\lambda(p) = 0$ for $\lambda,\sigma \in \{+1,-1\}$ always has two non-degenerate roots, as long as Eq.~\eqref{conditionrealroots} holds. Moreover, roots corresponding to $\sigma = +1$ can never coincide with roots for $\sigma = -1$ (for the same $\vec{p}$). The only degeneracy in the Lorentz-violating dispersion relations can thus come from roots for $\lambda = +1$ being equal to roots for $\lambda = -1$ (for the same value of $\sigma$). Inspection of Eqs.~\eqref{disprel1} and \eqref{disprel2} shows that this can only happen if $\delta(p) = 0$, which means that $k_{ZZ}^\mu \propto p^\mu$. In addition, for $\sigma = +1$ one also needs $p^2 = 0$ to solve the dispersion relation, which means that $k_{ZZ}^\mu$ must be lightlike, while for $\sigma = -1$, one needs $p^2 = m_Z^2$. Such that the degenerate momenta are:
\begin{equation}
p^\mu_{\rm degenerate} \left\{\begin{array}{ccc} = \pm \frac{m_Z k^\mu_{ZZ}}{\sqrt{k_{ZZ}^2}}, \quad k_{ZZ}^2 > 0 & {\rm for} & \sigma = +1 \\
\propto k^\mu_{ZZ}, \quad k_{ZZ}^2 = 0 & {\rm for} & \sigma = -1
\end{array}\right.\ .
\label{degeneratemomenta}
\end{equation}
In all these cases the Lorentz-violating term in the Lagrangian vanishes, which also happened in Ref.~\cite{ColladayMcDonaldNoordmansPotting}, where we showed that it did not result in serious problems for the quantization. We expect that this will also not happen in the current case.

\section{Branch-selection function}

In analogy with Ref.~\cite{ColladayMcDonaldNoordmansPotting},
we suspect that the sign of
\begin{equation}
\frac{\partial\Omega_\sigma^{\lambda}(p^0)}{\partial p^0}\ ,
\label{branchselec}
\end{equation}
evaluated at one of the roots, is observer Lorentz invariant and corresponds to the sign of the roots in concordant frames (observer frames where the Lorentz-violating coefficients can be considered small with respect to all other relevant quantities).
As pointed out in Ref.\ \cite{ColladayMcDonaldNoordmansPotting}, such a function can be used as an observer-Lorentz-invariant phase space factor and,
when used as the argument of a stepfunction under a four-dimensional momentum integral, as a function that selects the correct branch of the dispersion relation.
Since we have already shown that the two roots of $\Omega_\sigma^\lambda(p)$ -- for a particular value of $\sigma$ and $\lambda$ -- cannot coincide (for the same $\vec{p}$), it is now fairly easy to show that indeed the mentioned sign in invariant. First, notice that $\Omega_\sigma^\lambda(p) \rightarrow \infty$ for $p^0 \rightarrow \pm \infty$. So, plotted as a function of $p^0$, $\Omega_\sigma^\lambda(p)$ comes down from $+\infty$ (at large negative $p^0$) intersects the $p^0$-axis twice and goes up to $+\infty$ (at large positive $p^0$). Moreover, for timelike $k^\mu_{ZZ}$, $\Omega_\sigma^\lambda(p)$ is a continuous function. Consequently, at one root (the smallest), the derivative in Eq.~\eqref{branchselec} is negative, while at the other root (the larger one), the derivative is positive. Now, if $k^\mu_{ZZ}$ is spacelike, $\delta(p)$, and therefore $\Omega_\sigma^\lambda(p)$, can become imaginary. From Fig.~\ref{fig:disrel2plot} we see that the $p^0$ range where this happens lies in between the two roots of $\Omega_\sigma^\lambda(p)$. The conclusion it thus the same as for timelike $k_{ZZ}^\mu$. Moreover, this holds in any observer frame, since the argument did not assume any special frame. Finally, in concordant frames, the larger (smaller) root is positive (negative), showing that indeed, the sign of Eq.~\eqref{branchselec} corresponds to the sign of the roots in concordant frames. 

\section{Group velocity}
To address the group velocity, we define the four-vector $w_\sigma^{(\lambda)\mu} = \frac{\partial \Omega_\sigma^\lambda(p)}{\partial p_\mu}$. It is given by
\begin{eqnarray}
w_\sigma^{(\lambda)\mu} &=& 2\left[p^\mu + \lambda\left(\frac{1}{2}  + \sigma\frac{\tan^2(2\theta_w)\lambda\delta(p) - (\tfrac{1}{2}m_Z^2 - \lambda\delta(p))}{\sqrt{(m_Z^2 - 2\lambda\delta(p))^2+ 4\tan^2(2\theta_w)\delta(p)^2}}\right)\frac{\partial \delta(p)}{\partial p_\mu}\right] \notag \\
&=& 2\left[p^\mu + \frac{\lambda}{2}\left(\frac{p^2 - \tan^2(2\theta_w)\lambda\delta(p)}{p^2 - \tfrac{1}{2}m_Z^2 + \lambda \delta(p)}\right)\frac{\partial \delta(p))}{\partial p_\mu}\right]
\end{eqnarray}
where
\begin{equation}
\frac{\partial \delta(p)}{\partial p_\mu} = \frac{(p\cdot k_{ZZ})k_{ZZ}^\mu - k_{ZZ}^2 p^\mu}{\delta(p)}
\end{equation}
Since $w_\sigma^{(\lambda)0}$ corresponds to Eq.~\eqref{branchselec}, which has the same sign in any observer frame, we know that $w_\sigma^{(\lambda)\mu}$ must be timelike. Unfortunately, we cannot directly relate $w_\sigma^{(\lambda)\mu}$ to the group velocity, because $\Omega_\sigma^\lambda(p)$ is not a polynomial. However, $\Omega_T(p)\equiv \Omega_{+1}^+(p)\Omega_{-1}^+(p)\Omega_{+1}^-(p)\Omega_{-1}^-(p)$ is a polynomial, such that it can be written as
\begin{equation}
\Omega_T(p) = \prod_{i=1}^8(p^0 - \omega_i(\vec{p}))\ ,
\end{equation}
where $\omega_i(\vec{p})$ with $i\in \{1,\ldots,8\}$ are the eight non-degenerate roots of $\Omega_T(p)$ (except maybe at the momenta given in Eq.~\eqref{degeneratemomenta}). 
It follows that
\begin{subequations}
\begin{eqnarray}
\left.\frac{\partial\Omega_T(p)}{\partial p^0}\right|_{p^0 = \omega_j(\vec{p})} &=& \prod_{i\neq j}(\omega_j(\vec{p}) - \omega_i(\vec{p}))\ , \\
\left.\frac{\partial\Omega_T(p)}{\partial \vec{p}}\right|_{p^0 = \omega_j(\vec{p})} &=& -\frac{\partial \omega_j(\vec{p})}{\partial \vec{p}}\prod_{i\neq j}(\omega_j(\vec{p}) - \omega_i(\vec{p}))\ .
\end{eqnarray}
\end{subequations}
The first factor on the right-hand side of the second equation corresponds exactly to (minus) the group velocity (if we take $\omega_j(\vec{p})$ to be one of the concordant-frame positive roots). We have thus found that
\begin{equation}
\vec{v}_j = -\left[\frac{\partial\Omega_T(p)}{\partial \vec{p}} \middle/ \frac{\partial\Omega_T(p)}{\partial p^0}\right]_{p^0 = \omega_j(\vec{p})}\ .
\label{groupvel}
\end{equation}
On the other hand
\begin{equation}
\left.\frac{\partial\Omega_T(p)}{\partial p_\mu}\right|_{p^0 = \omega_\sigma^\lambda} = \left[\frac{\partial\Omega_\sigma^\lambda}{\partial p_\mu}\right]_{p^0 = \omega_\sigma^\lambda}\times \prod_{\sigma',\lambda'\neq \sigma,\lambda}\left[\Omega_{\sigma'}^{\lambda'}(p)\right]_{p^0 = \omega_\sigma^\lambda} = \left[w^{(\lambda)\mu}_\sigma\right]_{p^0 = \omega_\sigma^\lambda}\times \prod_{\sigma',\lambda'\neq \sigma,\lambda}\left[\Omega_{\sigma'}^{\lambda'}(p)\right]_{p^0 = \omega_\sigma^\lambda}\ ,
\end{equation}
where by $\omega_\sigma^\lambda$ we denote a root of $\Omega_\sigma^\lambda(p)$. The last factor on the right-hand side of the final equality is nonzero, since the roots are non-degenerate (except maybe at the momenta in Eq.~\eqref{degeneratemomenta}). Dividing the space components of this equation by the zeroth component, together with Eq.~\eqref{groupvel}, leads us to the conclusion that
\begin{equation}
\vec{v}_\sigma^\lambda = - \frac{\vec{w}^{(\lambda)}_\sigma}{w^{(\lambda)0}_\sigma}\ .
\end{equation}
Because $w^{(\lambda)\mu}_\sigma$ is timelike, the norm of the group velocity is thus always smaller than 1, which is necessary for causality of the theory.

\section{Extended Hamiltonian Formalism}

In order to elucidate the behavior of the particle states
derived above, we consider the classical mechanical limit that can
be obtained using an extended Hamiltonian formalism developed in \cite{extham}. 
The procedure involves picking an appropriate eigenvalue of the
off-shell dispersion relation as the extended Hamiltonian function.
Hamilton's equations can then be used to 
identify the appropriate classical wave packet velocity. 
A Legendre transformation then provides the appropriate Lagrangian
corresponding to that particular eigenvalue of the dispersion relation. 
In the present context, the appropriate Hamiltonian functions are given by
\begin{equation}
 \Hh_\sigma^{(\lambda)} = -\frac{e}{2 m} \Omega_\sigma^{(\lambda)}(p),
\end{equation}
where $e$ is a 1D metric (or einbein) on the world-line determined as 
a Lagrange-multiplier condition on the Lagrangian.
The physical four-velocity can be defined in terms of the four-vector
$w_\sigma^{(\lambda) \mu}$ in the previous section by using
the four-velocity of the corresponding wave packets which are
given by Hamilton's equations as
\begin{equation}
u^\mu = \frac{\partial \Hh_\sigma^{(\lambda)}}{\partial p_\mu} = 
- \frac{e}{2m} w_\sigma^{(\lambda) \mu}(p),
\end{equation}
The corresponding Lagrangians are found using a Legendre
transformation and are given as
\begin{equation}
\Ll_\sigma^{(\lambda)} = - u \cdot p - \Hh_\sigma^{(\lambda)}
= -\frac{e}{2 m}\left[ p^2 + \frac{1}{2} m_Z^2
\left( 1 - \frac{\sigma (m_Z^2 - 2 \lambda \delta(p))}
{\sqrt{(m_Z^2 - 2 \lambda \delta(p))^2+4 \delta(p)^2 \tan^2{2 \theta_W}}}
\right)\right]\ .
\end{equation}
This expression demonstrates that when $\delta(p)$ is small,
the solutions behave similarly to the conventional massive $Z$ and massless photon expressions,
while the mixing becomes significant when $\delta(p) \rightarrow m_Z^2 / 2$.  
In particular, the Lagrange densities of the $\lambda=+1$ states for
$\sigma = \pm 1$ become equal there indicating a symmetric behavior
between these polarizations of the photon and $Z$-boson states.
The Lagrangian can be expressed in terms of $u^\mu$ and $e$ by inverting
the velocity-momentum relation.
The expressions for $p(u)$ involve complicated solutions to a fourth-order polynomial
that can be formally written down, but they are of little obvious physical insight,
so  instead we work to second-order in $k$,
which is a good approximation at low energies where the mixing between
photon and $Z$-boson states is relatively small.
 
To second-order in $k$, these expressions are
\begin{equation}
 p^\mu \approx \frac{m}{e} u^\mu -  \left( \frac{\lambda (1 - \sigma)}{2} + 
 \frac{\sigma \delta_u(u) \tan^2{2 \theta_W}}{m e} \right)
 \frac{\partial \delta_u (u)}{\partial u_\mu},
\end{equation}
where $\delta_u(u) = \sqrt{(k \cdot u)^2 - k^2 u^2} $,
yielding the approximate Lagrangian
\begin{equation}
 \Ll_\sigma^{(\lambda)} \approx -\frac{m_Z}{2 e} u^2 + \frac{\lambda (1 - \sigma)}{2}\delta_u(u) 
 + \frac{\sigma}{2 m e} \delta_u(u)^2 \tan^2{2 \theta_W}
 - \frac{e(m^2 - \frac{1}{2} k^2 (1 - \sigma))}{4 m}(1 - \sigma).
\end{equation}
Note that this expression reduces to the same form as the massive,
CPT-violating Lagrangian in \cite{extham} when
$\theta_W \rightarrow 0$ and $\sigma = -1$, with a slightly different condition.
For the states $\sigma = -1$, the Lagrange multiplier condition
${\partial {\cal L} / \partial e}=0$ implies
\begin{equation}
 e^2 = \frac{1}{m^2 - k^2} \left( m^2 u^2 + \delta_u(u)^2 \tan^2{2 \theta_W} \right).
\end{equation}
When this value for $e$ is substituted into the Lagrangian, it becomes 
\begin{equation}
 \Ll \rightarrow - \sqrt{m^2 - k^2} \sqrt{u^2 + \frac{\delta_u(u)^2}{m^2} \tan^2{2 \theta_W}}
 + \lambda \delta_u(u).
\end{equation}
This provides a new example for a corresponding Finsler space function of the bipartite form 
analyzed in Ref.\ \cite{bipartite} when Wick-rotated to Euclidean space.
Interestingly, for the $\sigma = +1$ states,
the condition ${\partial {\cal L} / \partial e}=0 \rightarrow {\cal L} = 0$,
so this Lagrangian formalism fails to describe these states in this approximation.

\section{Discussion}
When considering the full $SU(2) \times U(1)$ electroweak sector of the Standard-Model Extension,
imposing gauge invariance forces the Lorentz- and CPT-breaking parameters to appear diagonally in the gauge-invariant fields.  
Electroweak symmetry breaking will induce mixing between these initial parameters which leads to
physical effects that depend on the energy scale. 
High-precision photon experiments place stringent bounds on the CPT-violating photon coupling, 
despite being corrected by a momentum-dependent term in Eq.\ (\ref{eff-kAF}).
Experiments performed at varying energies confirm that there can be no accidental cancellation of these
contributions, so the bounds quoted on $k_{AF}$ in the literature are safe from induced 
interference effects of and CPT violation in the $Z$-boson sector.
By extending the polarization vectors found in \cite{ColladayMcDonaldNoordmansPotting},
we were able to solve for the exact
dispersion relations of the full coupled system involving both
the $Z$ boson and the photon for the case of nonzero $k_{ZZ}$,
while $k_{AF}$ is assumed to be zero.
We find that, at low energies, two of the Lorentz-violating modes behave similarly to the massive $Z$ boson while the other two behave
as massless photons.
The photon states are always found to be spacelike, while the $Z$-boson
states are always timelike, which prevents Cherenkov-like $Z$-boson emission,
in contrast to what happens in the $W$ sector \cite{Wboson}.
The factor in Eq.\ (\ref{branchselec}) has been shown to be 
positive definite and can therefore be used as an phase-space normalization factor (as in Ref.\ \cite{ColladayMcDonaldNoordmansPotting, Wboson}), while the group velocities are always causal.
These facts are crucial to define the quantum theory consistently in nonconcordant frames where
the energies can go negative as described in \cite{ColladayMcDonaldNoordmansPotting} for 
the massive CPT-violating photon case.
The extended Hamiltonian formalism has been used to provide classical mechanical Lagrangians
for the particles involved. 
When working to second-order in $k$, in the perturbative regime, 
we find that the Lagrangian for the massive modes leads to a bipartite form,
while the massless modes lead to a trivial Lagrangian,
as happens in the conventional photon case.
The resulting nontrivial Lagrangian provides a new example of an physical model that can 
be described using bipartite Finsler geometry described in \cite{bipartite}.
The analysis done in this paper is complementary to the one in Ref.\ \cite{ColladayMcDonaldNoordmansPotting}
for the case of the photon with nonzero $k_{AF}$.
It would be interesting
to consider the general case in which both $k_{AF}$ and $k_{ZZ}$ are nonzero,
which should be a fully consistent model as well,
but one we expect to be quite challenging to analyze.
The results obtained in this work can be expected to be approximately
valid for nonzero, small $k_{AF}$, as long as the effects of the latter
on the process or quantity under consideration are negligible compared
to those of $k_{ZZ}$. 
For instance, when considering birefringence effects on the photon at low energy
this would mean that the first term on the right-hand side of
Eq.\ (\ref{eff-kAF}) is taken to be negligigle with respect to the second one.
Effects of nonzero $k_{AF}$ will be most pronounced at small four-momentum.
For instance, $p = 0$ will no longer be a solution of the photon
dispersion relation (see Eq.\ \eqref{spacetimelike} and below).

\vspace{6pt} 
\acknowledgments{R.~P.\ and J.~N.\ acknowledge support
by the Funda\c c\~ao para a Ci\^encia e a Tecnologia of Portugal (FCT)
through Projects No.\ UID/FIS/00099/2013 and No.\ SFRH/BPD/101403/2014
and Program No.\ POPH/FSE.
D.~C.\ acknowledges summer faculty development funds from New College of Florida.}



\begin{thebibliography}{999}
\bibitem{qgmodels1a}
  V.~A.~Kostelecky and S.~Samuel,
  Phys.\ Rev.\ D {\bf 39}, 683 (1989).
\bibitem{qgmodels1b} 
  V.~A.~Kostelecky and S.~Samuel,
  Phys.\ Rev.\ D {\bf 40}, 1886 (1989).
\bibitem{qgmodels2}
  V.~A.~Kostelecky and R.~Potting,
  Nucl.\ Phys.\ B {\bf 359}, 545 (1991).
\bibitem{qgmodels3}
  J.~R.~Ellis, N.~E.~Mavromatos and D.~V.~Nanopoulos,
  Gen.\ Rel.\ Grav.\  {\bf 31}, 1257 (1999).
\bibitem{qgmodels4} 
  R.~Gambini and J.~Pullin,
  Phys.\ Rev.\ D {\bf 59}, 124021 (1999).
\bibitem{qgmodels5} 
  C.~P.~Burgess, J.~M.~Cline, E.~Filotas, J.~Matias and G.~D.~Moore,
  JHEP {\bf 0203}, 043 (2002).
\bibitem{sme1}
  D.~Colladay and V.~A.~Kostelecky,
  Phys.\ Rev.\ D {\bf 55}, 6760 (1997).
\bibitem{sme2}
  D.~Colladay and V.~A.~Kostelecky,
  Phys.\ Rev.\ D {\bf 58}, 116002 (1998).
\bibitem{Gre02}
  O.~W.~Greenberg,
  Phys.\ Rev.\ Lett.\  {\bf 89}, 231602 (2002).
\bibitem{ColladayMcDonaldNoordmansPotting}
  D.~Colladay, P.~McDonald, J.~P.~Noordmans and R.~Potting,
  Phys.\ Rev.\ D {\bf 95}, 025025 (2017).
\bibitem{Wboson} 
  D.~Colladay, J.~P.~Noordmans and R.~Potting,
  Phys.\ Rev.\ D {\bf 96}, 035034 (2017).
\bibitem{CarrollFieldJackiw}
   S.~M.~Carroll, G.~B.~Field and R.~Jackiw,
  Phys.\ Rev.\ D {\bf 41}, 1231 (1990).
\bibitem{datatables} 
  V.~A.~Kostelecky and N.~Russell,
  Rev.\ Mod.\ Phys.\  {\bf 83}, 11 (2011)
  (updated version: arXiv:0801.0287 [hep-ph]).
\bibitem{KosteleckyMewes1}
  V.~A.~Kostelecky and M.~Mewes,
  Phys.\ Rev.\ D {\bf 80}, 015020 (2009).
\bibitem{extham}
  D.~Colladay,
  Phys.\ Lett.\ B {\bf 772}, 694 (2017).
\bibitem{bipartite}
  V.~Alan Kosteleck\'y, N.~Russell and R.~Tso,
  Phys.\ Lett.\ B {\bf 716}, 470 (2012).

  
\end{thebibliography}
\end{document}